## Title
- Full title: Kinetic approach to superconductivity hidden behind a competing order
- Short title: Thermal-quench-induced superconductivity

## Authors
H. Oike,[1,2]* M. Kamitani, Y. Tokura,[1,2] F. Kagawa,[1,2]*

## Affiliations
[1]RIKEN Center for Emergent Matter Science (CEMS), Wako 351-0198, Japan
[2]Department of Applied Physics, The University of Tokyo, Tokyo 113-8656, Japan
*Correspondence to: oike@ap.t.u-tokyo.ac.jp; kagawa@ap.t.u-tokyo.ac.jp

## Abstract
Exploration for superconductivity is one of the research frontiers in condensed matter physics. In strongly correlated electron systems, the emergence of superconductivity is often inhibited by the formation of a thermodynamically more stable magnetic/charge order. Thus, to develop the superconductivity as the thermodynamically most stable state, the free-energy balance between the superconductivity and the competing order has been controlled mainly by changing thermodynamic parameters, such as the physical/chemical pressure and carrier density. However, such a thermodynamic approach may not be the only way to materialize the superconductivity. Here, we present a new kinetic approach to avoiding the competing order and thereby inducing persistent superconductivity. In the transition-metal dichalcogenide $IrTe_2$ as an example, by utilizing current-pulse-based rapid cooling up to ~$10^7$ K s$^{-1}$, we successfully kinetically avoid a first-order phase transition to a competing charge order and uncover metastable superconductivity hidden behind. Because the electronic states at low temperatures depend on the history of thermal quenching, electric pulse applications enable non-volatile and reversible switching of the metastable superconductivity, a unique advantage of the kinetic approach. Thus, our findings provide a new approach to developing and manipulating superconductivity beyond the framework of thermodynamics.

## MAIN TEXT

### Introduction

Thermodynamics states that materials at equilibrium adopt the lowest free-energy state when several states are competing; this principle has provided solid and universal guidance in controlling static phases of matter. For instance, in alloys, one can materialize an intended structural form out of various candidate forms by tuning their composition, a control parameter of their free-energy balance (*1*). Similarly, in strongly correlated electron systems, the internal degrees of freedom of electron can exhibit various ordered phases at low temperatures, and their free-energy balance is closely associated with electron bandwidth and band filling. Thus, pressure and/or carrier doping have been exclusively exploited for searching superconductivity as the ground state (*2-4*). This approach is schematically illustrated in Fig. 1, which we call the thermodynamic approach.

Although the thermodynamic approach is literally based on thermodynamics and can hence be applied to a variety of systems, there is another phase-control method that is not properly described within the framework of thermodynamics, that is, the kinetic approach, a main subject in this report. The kinetic approach is based on the fact that within a considered time range, a system does not necessarily trace the global free-energy minimum and may be trapped in a local minimum, a metastable state. For instance, carbon steel changes its form upon cooling from face-centered-cubic to body-centered-cubic according to the temperature-dependent minimum free-energy (*5*). However, this phase change is a first-order transition initiated by nucleation, and it can therefore be kinetically avoided when the system is sufficiently rapidly cooled to a certain low temperature. Such a kinetic approach based on thermal quenching has long been employed in the fields of ferrous metallurgy and supercooled liquids to materialize higher-energy metastable structural forms, such as martensite (*6*) and glass (*7*, *8*); however, only recently has thermal quenching far beyond standard cooling rates come into use for controlling electronic/magnetic states in condensed matters (*9−11*). Here, we find that thermal quenching is capable of materializing persistent superconductivity that is otherwise superseded by a thermodynamically more stable magnetic/charge order. Such a kinetic approach enables nonvolatile and reversible switching of superconductivity with pulse application, providing a basis for superconducting devices that do not require band-filling control through gate electrodes or band-width control through piezo elements.

## Results
### Material selection

For selecting a candidate material, we set the two criteria depicted in Fig. 1. First, the ground state is a magnetic/charge order, but a superconducting phase neighbors it with physical/chemical pressure or carrier doping as a control parameter. Second, the magnetic/charge order is formed through a first-order transition. We envisaged that in such a material, superconductivity would emerge as a metastable state if the formation of the more stable competing order is avoided kinetically under rapid cooling (see "Kinetic approach" in Fig. 1). Among many candidate materials, we targeted the transition-metal dichalcogenide $IrTe_2$, the phase diagram of which explicitly shows that a superconducting phase with $T_c^{onset} \approx 3.0$–$3.1$ K appears when a first-order charge-ordering transition is thermodynamically suppressed, in this case by element substitution (*12−14*) (Fig. 2A). A recent scanning tunneling microscopy study on pristine $IrTe_2$ has shown that rapid cooling results in the emergence of superconducting patches at the surface (*15*); however, the signature of superconductivity has never been detected in macroscopic physical properties, indicating that the validity of our hypothetical kinetic approach toward macroscopic superconductivity still remains unclear.

### Thermal quenching

To this end, we focused on standard electric resistivity measurements and performed thermal quenching at an unconventionally high rate, $>10^6$–$10^7$ K s$^{-1}$, from a sufficiently high temperature ($\approx 300$–$400$ K) to 4 K. To achieve this quenching, we applied electric-pulse heating to a sample in contact with a substrate at 4 K (Fig. 2B), which is inevitably followed by rapid cooling after the pulse ends because of a substantial thermal gradient between the sample and the substrate (Fig. 2C) (*16*). In implementing the kinetic approach based on thermal quenching, we also considered the sample volume *V*: Just as rapid cooling invariably results in a lower nucleation probability in terms of time, sample miniaturization makes nucleation less probable in terms of the number of nucleation sites contained in a specimen (*17*, *18*). Thus, sample miniaturization, in principle, can supplement the kinetic approach. Nevertheless, the sample volume is a "static" state variable and can also thermodynamically affect the free-energy balance through the change in the ratio between the bulk and surface energies. To exclusively highlight the kinetic aspect of the approach, we therefore focused on how the low-temperature electronic state varies when the quenching to 4 K is applied to IrTe$_2$ thin plates with different volumes.

The results are displayed in Fig. 2D−G. In a relatively large thin plate, #1 ($V \approx 130$ μm$^3$), a sharp resistivity increase signaling the charge-order formation was observed at $T_{\mathrm{CO}}^{\mathrm{cool}} \approx 250$ K under slow cooling (Fig. 2D), and accordingly, the temperature-resistivity profile exhibited no signature of superconductivity at the lowest temperature (Fig. 2E). This behavior is in good agreement with that of the bulk sample (*19*), and it can be regarded as a reference when the charge order is fully formed in a considered specimen. Notably, after performing quenching to 4 K at a cooling rate of ~10$^5$ K s$^{-1}$ by utilizing the electric-pulse heating (Figs. S1−S3), a signature of superconductivity with $T_{\mathrm{c}}^{\mathrm{onset}} \approx 3.1$ K emerged (Fig. 2E), consistent with our working hypothesis. However, zero resistance was not reached, probably because of an insufficient superconducting volume fraction.

To achieve quenching-induced zero resistance with an experimentally accessible quenching rate, we next targeted a smaller sample: #2 ($V \approx 15$ μm$^3$). As expected, a low-temperature resistivity profile resembling that of sample #1 post-quenching was obtained without implementing quenching (Fig. 2G), consistent with the idea that sample miniaturization can supplement the kinetic approach although a possible change in the free-energy balance cannot be ruled out. At higher temperatures, a broadened, relatively weak increase in resistivity was observed at $T_{\mathrm{CO}}^{\mathrm{cool}} \approx 200$ K (Fig. 2F), which is considerably lower than the case of sample #1. These observations suggest that the charge ordering was avoided in a certain fraction even under slow cooling, thus leading to superconductivity as a minor phase. Remarkably, we found that when the quenching (~10$^7$ K s$^{-1}$) is applied, the quenched state exhibits a relatively sharp drop with $T_{\mathrm{c}}^{\mathrm{onset}} \approx 3.4$ K and reaches zero resistance (Fig. 2G). Similar behavior was also observed in another thin plate with a comparable volume, #3 ($V \approx 18$ μm$^3$) (see Fig. S4), thus establishing that zero resistance is achieved by quenching. The $T_{\mathrm{c}}^{\mathrm{onset}}$ of the emergent superconductivity in the quenched states is plotted in the phase diagram and found to be comparable to, or slightly higher than, those of doped IrTe$_2$ (Fig. 2A).

**Quenching-rate dependence**

To gain further insight into how the applied quenching rate correlates with the emergence of the superconductivity, we tailored the quenching rate with ramp pulses (Fig. S5) and systematically examined the low-temperature resistivity post-quenching for thin plate #3. The results are summarized as a resistivity contour plot in Fig. 3. Here, two important aspects can be highlighted. First, as the rate of quenching to 4 K increases, the zero-resistance temperature $T_{\mathrm{c}}^{\mathrm{zero}}$ continuously increases, whereas $T_{\mathrm{c}}^{\mathrm{onset}}$ varies only weakly. Although the sample smallness makes it difficult to estimate the superconducting volume fraction via magnetic susceptibility measurements, this behavior can be understood by considering that the quenching rate continuously changes the superconducting volume fraction. Second, the systematic variations indicate that the quenched state hosting superconductivity is metastable. Moreover, the zero-resistance state is found to disappear if the quenched state is heated above 50 K and subsequently cooled (see Fig. S6), consistent with the general propensity that a quenched state is metastable and it reverts to a slowly cooled state if an annealing procedure is properly performed. Conversely, the metastability of the quenched state is practically robust below 50 K (greater than 1 week at 10 K), and thus, a magnetic-field-induced superconducting-to-normal transition shows a reversible behavior (Fig. S7).

**Non-volatile switching**

Finally, we show that the bistability consisting of the nonzero- and zero-resistance states, unveiled by the kinetic approach, raises an interesting possibility, that is, the design of a reversible superconducting switch based on applying electric pulses. The working principle considered here is illustrated in the schematic (Fig. 4A), and it is analogous to that of conventional phase-change memories (*20*, *21*), in which the operation is ultimately based on rapid temperature control using electric/optical pulses. In the forward switching from the nonzero-resistance slowly cooled to zero-resistance quenched states (the 'SET' process), we used electric-pulse heating followed by rapid cooling at ~$10^7$ K s$^{-1}$ to a sample-holder temperature below $T_c^{zero}$. Similarly, the described annealing process of the quenched state can be substituted by the application of an electric pulse with relatively weak magnitude: during the pulse application (≈2 s), the quenched state is held at temperatures above 50 K, and after the pulse ends, the sample is promptly cooled (~$10^7$ K s$^{-1}$) to the sample-holder temperature, completing the backward switching (the 'RESET' process).

The typical resistance switching behavior is displayed in Fig. 4B, C. Current-pulse-induced switching between zero- and nonzero-resistance states was successfully observed at 2.4 K in a non-volatile manner without changing the sample-holder temperature. The switching is reproducible (Fig. 4D), thereby highlighting the deterministic nature of the working principle described (Fig. 4A). Such superconducting phase control with pulse application is a unique advantage of the kinetic approach implemented by thermal quenching. If the kinetic approach is implemented only by the sample miniaturization, superconductivity invariably appears even under slow cooling (*22*), thus losing its controllability. Moreover, if one seeks to design a superconducting switching function within the framework of the thermodynamic approach, the device would have to be equipped with certain elements that enable in-situ and reversible control of pressure or carrier density at low temperatures (*23–25*).

**Discussion**

When the emergence of superconductivity is inhibited by a magnetic/charge order, it has widely been considered that, for the superconductivity to develop, the free-energy balance has to be modulated to make the superconducting state energetically most favored. To circumvent this thermodynamic constraint, highly nonequilibrium methods using intense light pulses have recently been exploited, and the superconductivity is actually induced out of a certain competing phase but only transiently (*26*). Our observations have shown that when the overall feature of a thermo-equilibrium phase diagram agrees with that shown in Fig. 1, persistent superconductivity can be realized by another nonequilibrium method, that is, quenching the system to low temperatures such that nucleation of the competing order is no longer expected in the considered time range. Thus, our work provides some of the materials that have been categorized as non-superconducting with a fresh chance to yield metastable superconducting behavior.

## Materials and Methods

### Sample preparation

Bulk single crystals of $IrTe_2$ were synthesized using the Te-flux method according to the literature (*19*). Sub-micrometer-thick crystals were exfoliated from a bulk single crystal with Scotch tape and transferred onto silicon (Si) or polyethylene naphthalate (PEN) substrates. Gold electrodes were prepared on the substrate and sample surface using photolithography methods. The electrodes on the sample surface were connected with those on the substrate by tungsten deposition using a focused-ion beam to ensure the electrical connection between them. An image of a typical sample is shown in Fig. 2B in the main text. The contact resistance depends on the area of the current electrode, and it is typically 1–3 Ω for each contact. The sample thickness was measured using scanning electron microscopy.

### Resistivity measurements

The resistivity under slow cooling was measured with the conventional four-probe method. A load resistor of 10 kΩ was connected in series with the sample. An AC voltage excitation of 333 Hz with a magnitude corresponding to 10 µA was generated at a lock-in amplifier (Signal recovery, 7270) and applied to the circuit. Signals from the voltage probes were amplified with a low-noise preamplifier (NF Corporation, SA-410F3 or Stanford Research, SR560) and measured with the lock-in amplifier. The current flowing through the circuit was measured by probing the voltage drop at the load resistor with another lock-in amplifier.

### Pulse application

To feed a current pulse with a sufficient magnitude (160 and 19 mA for samples on Si and PEN substrates, respectively), a trapezoidal voltage that was generated by a function generator (NF Corporation, WF1947) was amplified by a high-speed bipolar power supply (NF Corporation, HSA4014). A load resistor of 2.5 Ω was connected in series with the sample and used to calculate the current flowing through the circuit. The time-varying voltages at the load resistor and the sample voltage-probes were amplified by a differential amplifier (NF Corporation, 5307) and a low-noise preamplifier (NF Corporation, SA-410F3 or Stanford Research, SR560), respectively, and the amplified voltages were monitored using a digitizer (National Instruments, NI PXIe-5122) or an analog input module (National Instruments, NI 9239). Thus, we obtained the time profiles of the current and sample resistivity during the pulse application. A switch system (Keithley 7001 equipped with 7011S) was used to switch the circuit between the resistivity-measurement and pulse-application setups.

# H2: Supplementary Materials

Supplementary Materials and Methods
Fig. S1. Sample heating by a trapezoidal current-pulse application, followed by rapid cooling.
Fig. S2. Measurements of the highest quenching rate achieved in the present experiments.
Fig. S3. Simulation results of the quenching process for an $IrTe_2$ thin plate on Si and PEN substrates.
Fig. S4. Creation of persistent superconductivity by a current application in the $IrTe_2$ thin plate #3.
Fig. S5. Manipulation of the quenching rate by controlling the pulse fall time.
Fig. S6. Annealing process of the quenched metastable state in the $IrTe_2$ thin plate #3 on the Si substrate.
Fig. S7. Magnetic-field effects on the emergent superconductivity in the quenched metastable states.

## Acknowledgments

**General**: F.K. and H.O. thank Y. Kawasugi and R. Takagi for their valuable discussions.

**Funding:** This work was partially supported by JSPS KAKENHI (Grant Nos. 18K13512 and 18H01168).

**Author contributions:** H.O. performed all the experiments and analyzed the data. M.K. grew the single crystals used for the study. F.K. and H.O. planned the project. H.O. and F.K. wrote the manuscript. H.O., F.K., and Y.T. discussed the results and commented on the manuscript.

**Competing interests:** The authors declare no competing financial interests.

**Figures and Tables**

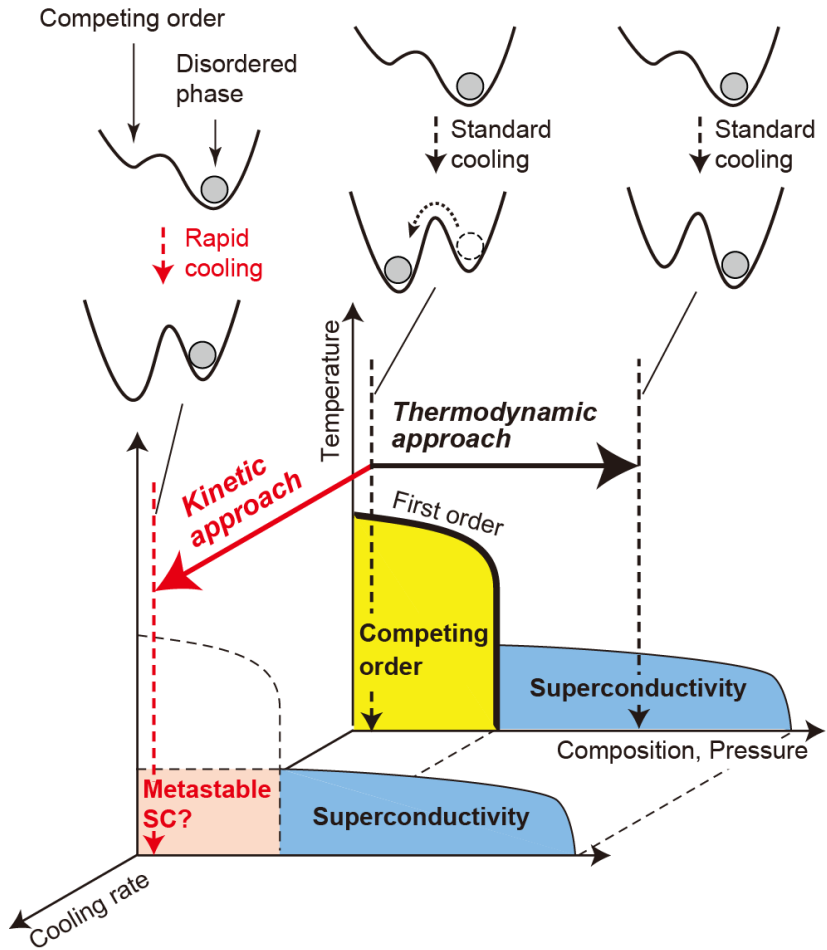

**Fig. 1. Scheme for thermodynamic and kinetic approaches to realizing superconductivity in certain strongly correlated electron systems.** The conceptual electronic phase diagram considered in this study is displayed with pressure/carrier doping as a control parameter. The double-well and ball represent temperature-dependent schematic free-energy landscapes and realized electronic states in each cooling process, respectively. The thermodynamic approach, which can be advanced by increasing pressure or carrier doping, results in a change in the lowest free-energy state, from a certain competing order to a superconducting state. By contrast, the kinetic approach, which can be advanced by rapid cooling, allows the system to kinetically avoid the first-order phase transition to the competing order and thus to remain in a metastable supercooled state, which is expected to eventually turn into superconducting at low temperatures.

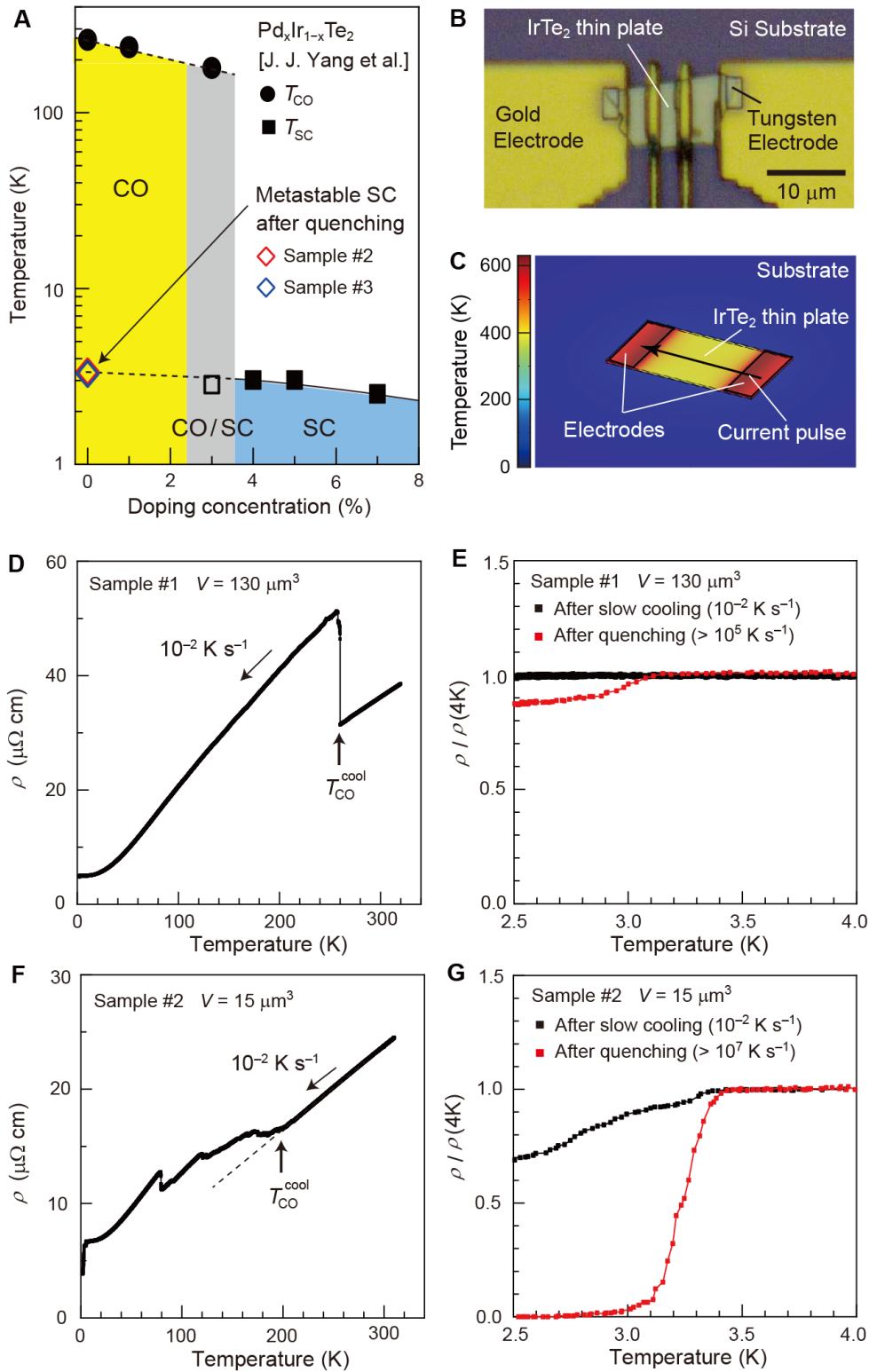

**Fig. 2. Demonstration of the kinetic approach to a superconducting state in non-doped IrTe₂.** (**A**) Electronic phase diagram of Pd-doped IrTe$_2$. CO and SC denote the charge-ordered and superconducting phases, respectively. The open diamonds represent the onset temperature of the emergent superconductivity in the quenched metastable state of non-doped IrTe$_2$ samples: #2 and #3. The data for the doped materials are taken from ref. (*12*). (**B**) A photograph of a submicrometer-thick IrTe$_2$. (**C**) A schematic of the current-pulse-based rapid-cooling method used in this study. At the end of an electric pulse, a large thermal gradient is realized between the sample and the substrate, thus enabling the rapid cooling of the sample after the pulse ends. The schematic is a result of our numerical simulation (*16*). (**D, F**) Overall temperature-resistivity profiles of samples #1 (D) and #2 (F), the volume of which is ≈130 and 15 $\mu m^3$, respectively. (**E, G**) The temperature-resistivity profiles of samples #1 (E) and #2 (G) near the superconductivity-onset temperature, measured after being slowly cooled and thermally quenched to 4 K.

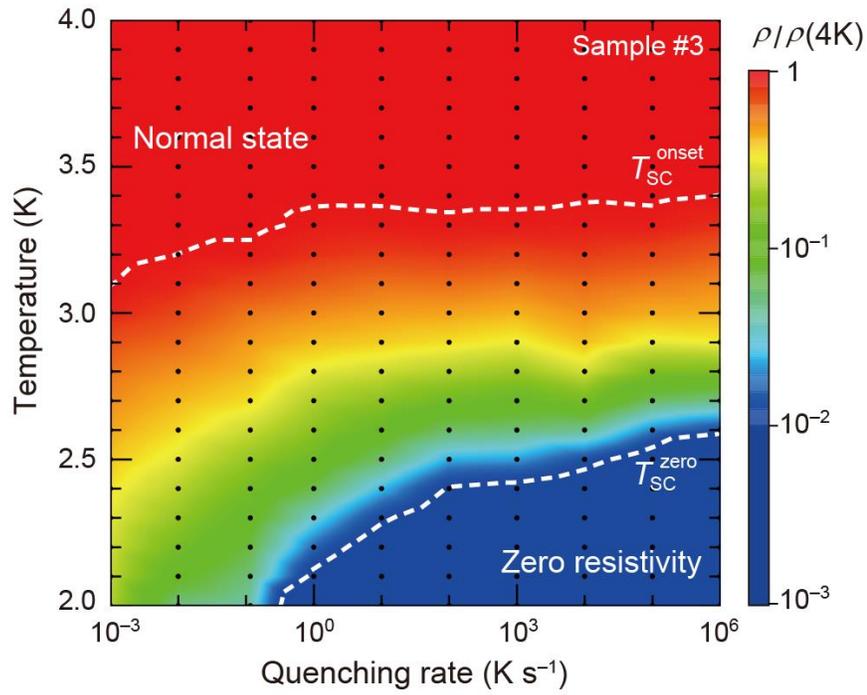

**Fig. 3. Contour plot of the electric resistivity post-quenching to 4 K at various quenching rates in sample #3.** This figure represents an interplay between the quenching rate and the emergent superconductivity in non-doped $IrTe_2$. To clearly show the superconducting-transition onset, the data are normalized by the value at 4 K post-quenching at each quenching rate.

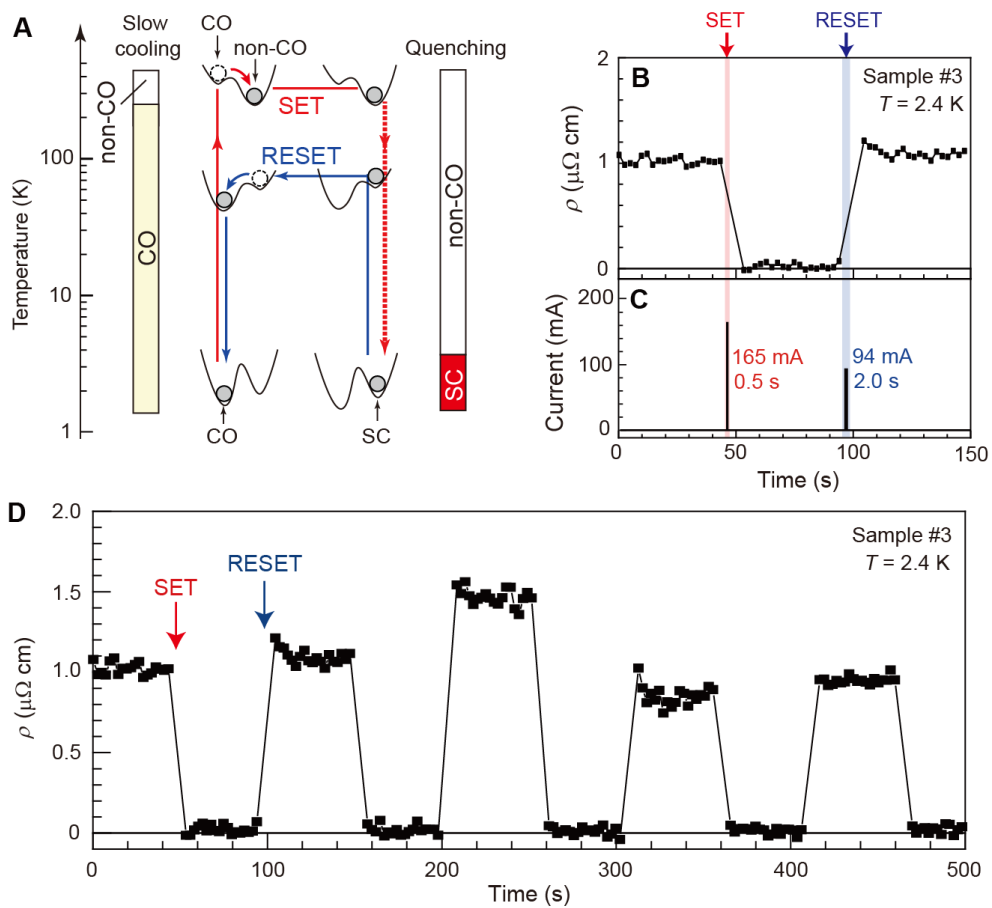

**Fig. 4. Non-volatile superconducting phase-change by the application of current pulses.** (**A**) The scheme for repeatable switching between the superconducting (SC) and charge-ordered (CO) states in terms of a schematic free-energy landscape and the kinetic approach. In the SET process, the application of a high-intensity current pulse results in the rapid cooling of the sample from a temperature above the charge ordering. In the RESET process, a pulse with moderate intensity and longer width is applied to heat the quenched state to a certain temperature below the charge ordering, facilitating relaxation to the charge order, which is more stable than the quenched metastable state. (**B, C**) The single-cycle operation of the non-volatiles switching. The time profiles of the resistivity and current are shown in (B) and (C), respectively. (**D**) Repetitive switching between the zero-resistance and nonzero-resistance states by the application of current pulses.

## Supplementary Materials

### Supplementary Materials and Methods

#### Quenching method

As a typical example of the electric-pulse rapid heating followed by rapid cooling, here we discuss the time profiles of the current and resistivity for the case that the quenching rate is not sufficiently high to kinetically avoid the formation of the charge order. Fig. S1A,B displays the results of the sample #4 ($V \approx 180$ µm$^3$) on PEN substrate, in which the pulse amplitude was set at 19 mA and the rise, top-duration and fall times were set at 1, 1 and 3 s, respectively. The time-varying resistivity exhibits a complex profile because the melting and formation of the charge order are superimposed on the electric-pulse heating and subsequent cooling processes, respectively (Fig. S1B). However, when the time-varying resistivity is plotted as a function of current (Fig. S1C), the current-resistivity profile exhibits qualitatively similar behavior to the temperature-resistivity profile measured at a low temperature-sweep rate, $10^{-2}$ K s$^{-1}$ (Fig. S1D), indicating that the pulse application results in a rapid sweeping of the sample temperature, in this exemplary case, on the order of $10^2$ K s$^{-1}$.

Given this, below, we explain the time-resistivity profile (Fig. S1B) in greater detail. Before the pulse application, sample #4 on the PEN substrate was slowly cooled from 320 to 30 K, and thus, the initial state is the charge-ordered state (Fig. S1D). To measure the resistivity of the electronic states before and after the pulse application, the pulse is offset by a low DC current, the value of which is one order of magnitude smaller than that of the pulse top (Fig. S1A). During the pulse rising, the resistivity increases because of the Joule heating of the sample; it subsequently exhibits a sharp drop due to the melting of the charge-order. During the pulse top, the resistivity remained constant, indicating that the Joule heating is balanced with the thermal dissipation from the sample to the low-temperature substrate (see also Fig. S3B). During the pulse falling, the Joule heating decreases accordingly; thus, in the course of the resistivity decrease, a sharp resistivity increase is observed due to the re-formation of the charge order. At the end of the fall time, the resistivity returns to the initial value, thus indicting that despite the intense Joule heating, the pulse application causes no appreciable damage to the sample.

#### Control of quenching rate

To control the quenching rate, we manipulated the fall time of the pulse. Fig. S5A–D shows the time-varying current and resistivity during the pulse falling with various fall times. Here, we set the origin of the time at the beginning of the pulse fall, namely, at the moment when the sample temperature starts to decrease. The time profiles of the current and resistivity show a good correspondence, thus demonstrating that the quenching rate can be systematically tailored by manipulating the fall time. Whereas sample #4 ($V \approx 180$ µm$^3$) exhibits a peak structure in the time-resistivity profile because of the charge ordering upon cooling (Figs. S1B and S5D), such a feature is not discernible in sample #3 ($V \approx 18$ µm$^3$) because the full development of the charge ordering is kinetically avoided by the combination of the sample miniaturization and rapid cooling. The highest cooling rate is achieved when the fall time is set to "zero" within the resolution of the function generator, and it is higher than $10^7$ and $10^5$ K s$^{-1}$ for samples #3 (on the Si substrate) and #4 (on the PEN substrate), respectively (Fig. S2). The difference in the highest cooling rate originates from that in thermal conductivity between Si and PEN substrates.

Numerical simulation

To examine whether the orders of the highest cooling rates estimated above are reasonable, we performed a finite element method using commercial software (COMSOL Multiphysics) and simulated the quenching procedure for an IrTe$_2$ thin plate on Si and PEN substrates. In the numerical simulation, we employed the geometry shown in Fig. S3A,B, in which the IrTe$_2$ sample is modeled as a rectangular thin plate with dimensions of $16\times7\times0.2$ μm$^3$ and the substrate dimensions are $1\times1\times0.5$ mm$^3$. We set values of thermal conductivity $\kappa$, specific heat $C_p$, and resistivity $\rho$ as follows:

$\kappa = 10$ W m$^{-1}$ s$^{-1}$, $C_p = 2\times10^6$ J K$^{-1}$ m$^{-3}$ and $\rho = 50$ μΩ cm for IrTe$_2$;

$\kappa = 150$ W m$^{-1}$ s$^{-1}$, $C_p = 2\times10^5$ J K$^{-1}$ m$^{-3}$ for Si substrate;

$\kappa = 0.3$ W m$^{-1}$ s$^{-1}$, $C_p = 2\times10^5$ J K$^{-1}$ m$^{-3}$ for PEN substrate;

and the contact resistance of the current electrodes was set at 1.5 Ω.

In reality, these parameters should vary with temperature, but for simplicity, we treated them as constants and performed an order-of-estimate calculation of the quenching rate. To model heat transfer with a thermal impedance from the sample to the substrate, we also assumed a contact layer between them. The thermal conductivity and heat capacity of the contact layer were treated as adjustable parameters, and they were determined to reproduce the experimental observations that the sample is heated to ~400 K by current applications of 160 and 19 mA for an IrTe$_2$ thin plate on Si and PEN substrates, respectively. In the simulation, the trapezoidal current pulse shown in Fig. S3C was applied to the samples, with the same magnitude of current as that used in the experiments. Although the sample temperature is not uniform, we extracted the temperature at the center of the sample and plotted its time profile in Fig. S3D. The quenching rate obtained in the simulation is $4\times10^7$ and $2\times10^6$ K s$^{-1}$ for the sample on the Si and PEN substrates, respectively, and it is nearly the same order as that estimated from the experiments (Fig. S2C,D). Thus, the numerical simulation supports the high quenching rate estimated from the experiment.

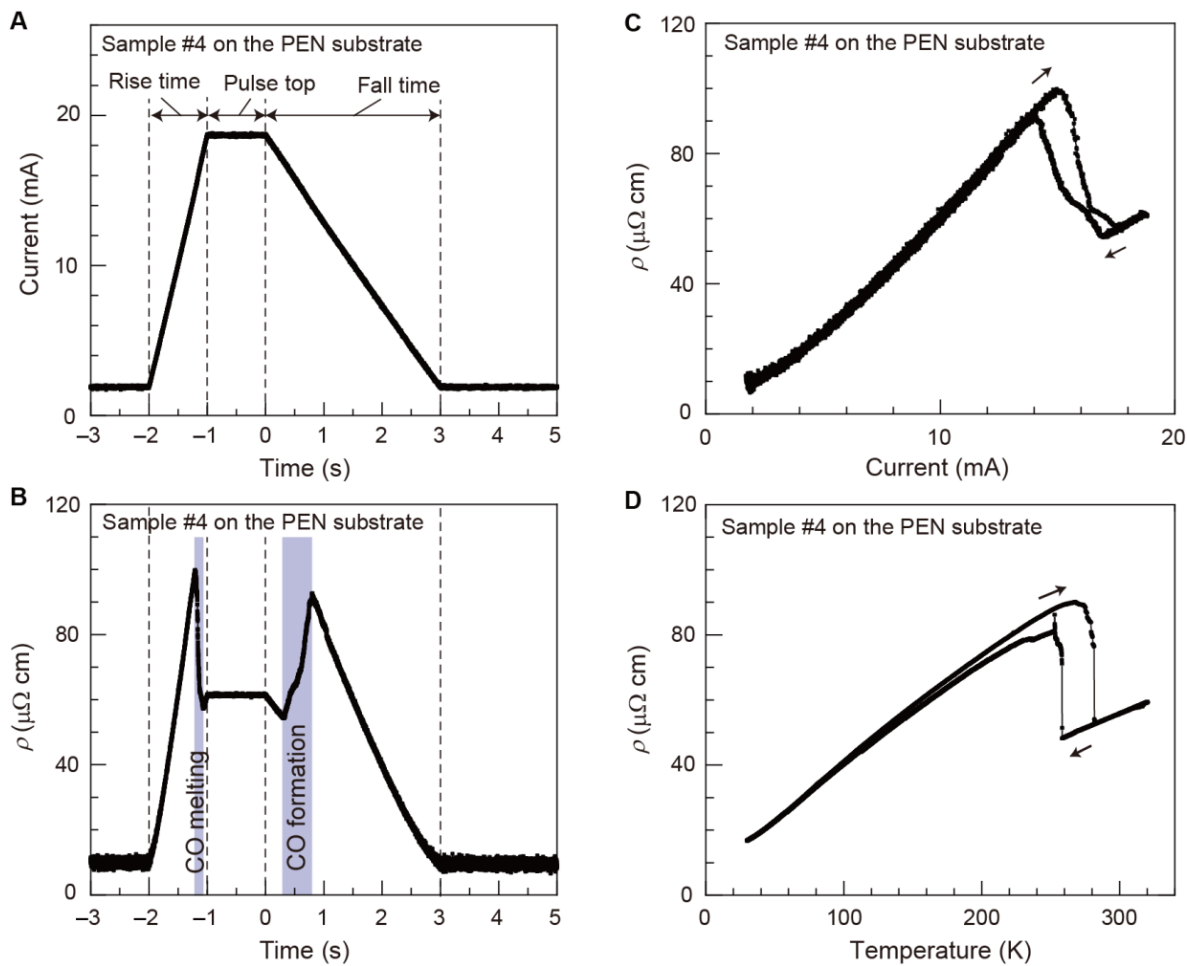

**Fig. S1. Sample heating by a trapezoidal current-pulse application, followed by rapid cooling.** (**A** and **B**) The time profiles of the applied current pulse (A) and the corresponding change in resistivity (B) in the IrTe$_2$ thin plate #4 ($V \approx 180$ μm$^3$) on PEN substrate. The vertical broken lines are visual guides, therein delimiting the time boundary between the pulse rising, pulse top, and pulse falling. CO in (B) denotes the charge order. The origin of the time is set at the beginning of the pulse fall. (**C**) The current-resistivity profile during the pulse application, extracted from (A) and (B). (**D**) The temperature-resistivity profile of the IrTe$_2$ thin plate #4, measured at a temperature-sweeping rate of $10^{-2}$ K s$^{-1}$.

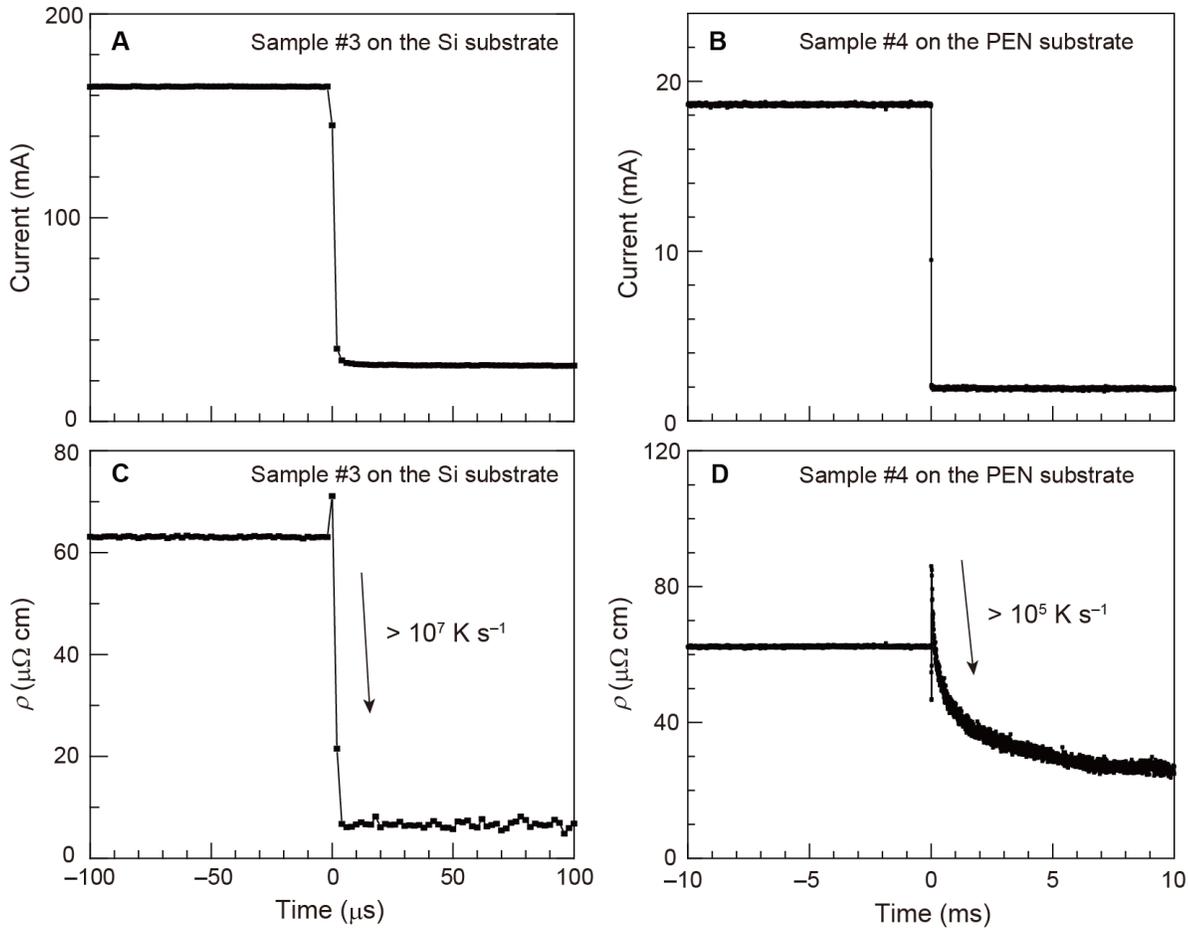

**Fig. S2. Measurements of the highest quenching rate achieved in the present experiments.**
(**A** to **D**) The time profiles of the current applied to the IrTe$_2$ thin plate #3 ($V \approx 18$ μm$^3$) on the Si substrate (A) and #4 ($V \approx 180$ μm$^3$) on the PEN substrate (B) and the corresponding change in resistivity for sample #3 (C) and #4 (D). The origin of the time is set at the beginning of the pulse fall. From the time-resistivity profiles, the order of magnitudes of the achieved quenching rate is estimated as shown in (C) and (D).

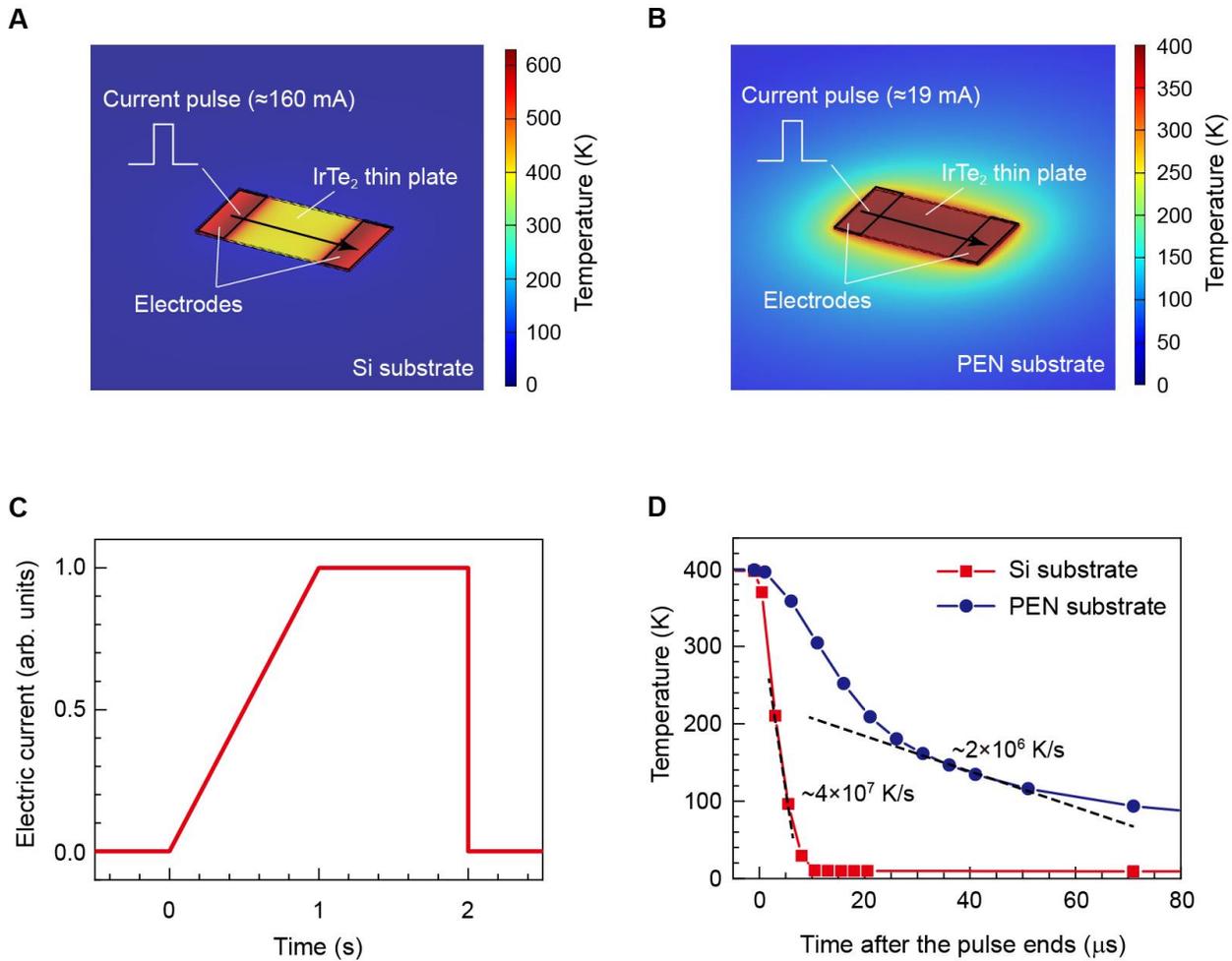

**Fig. S3. Simulation results of the quenching process for an IrTe$_2$ thin plate on Si and PEN substrates.** (**A** and **B**) The spatial distribution of temperature during the pulse top for a IrTe$_2$ thin plate on Si substrate (A) and that on a PEN substrate (B). Before the pulse application, the sample and substrate are in equilibrium at 4 K. (**C**) The time profile of the current amplitude used in the simulations. The value of the pulse top is 160 and 19 mA for a thin plate on the Si and PEN substrates, respectively. (**D**) The time profiles of the sample temperature during the quenching following the pulse application. The average cooling rates in 100–200 K are estimated from the profiles.

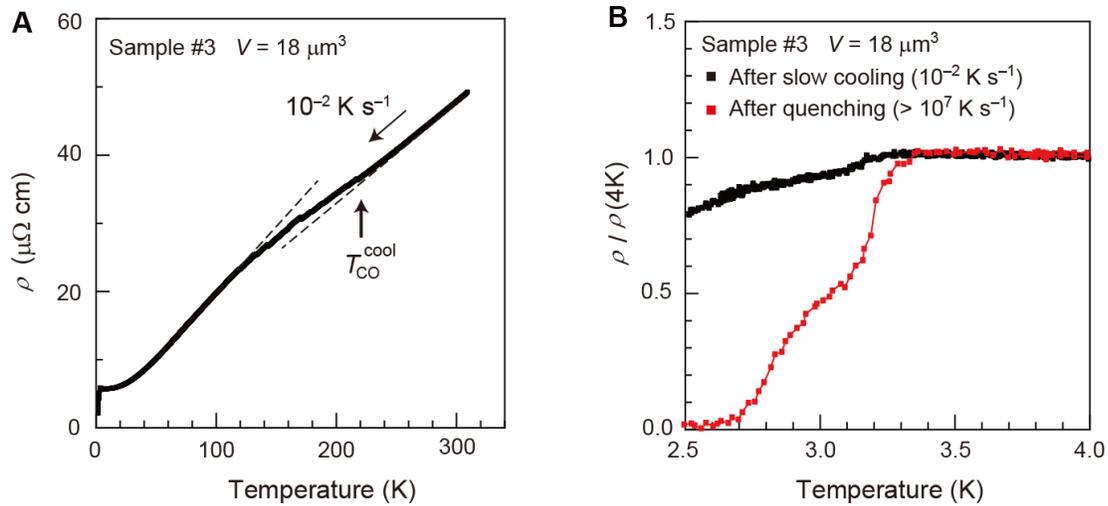

**Fig. S4. Creation of persistent superconductivity by a current application in the IrTe$_2$ thin plate #3.** (**A**) The temperature-resistivity profile measured at a cooling rate of $10^{-2}$ K s$^{-1}$. A weak resistivity anomaly accompanying the charge order is observed, the onset of which is indicated by the arrow. (**B**) The temperature-resistivity profiles at low temperatures after slow cooling and quenching to 4 K.

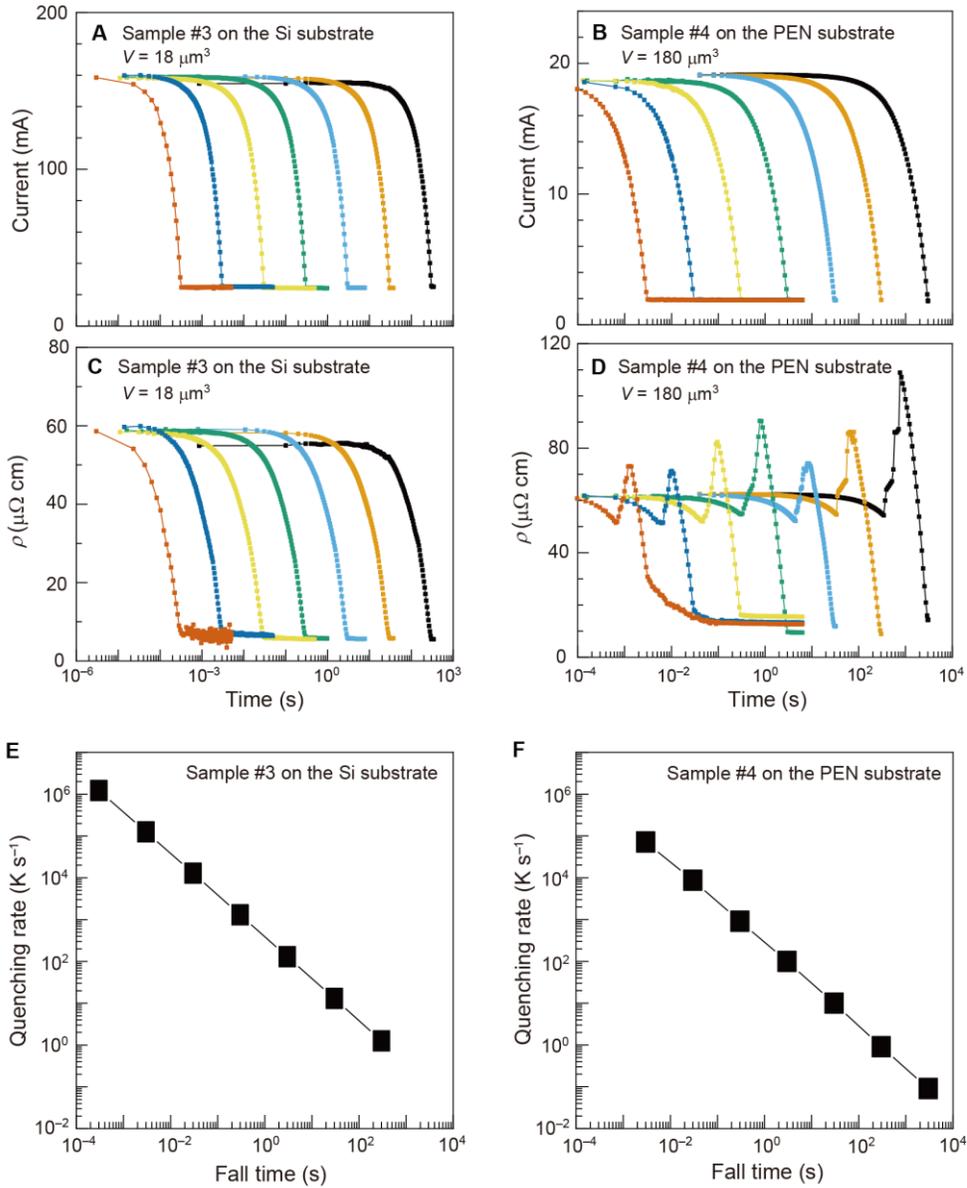

**Fig. S5. Manipulation of the quenching rate by controlling the pulse fall time.** (**A** to **D**) The time profiles of the current applied to the IrTe$_2$ thin plate #3 on the Si substrate (A) and #4 on the PEN substrate (B), as well as the corresponding change in resistivity for #3 (C) and #4 (D). The time profiles displayed in (C) and (D) were recorded under the current profile with the corresponding color in (A) and (B), respectively. The origin of the time is set at the beginning of the pulse fall. (**E** and **F**) Estimated quenching rate at each fall time for thin plates #3 (E) and #4 (F).

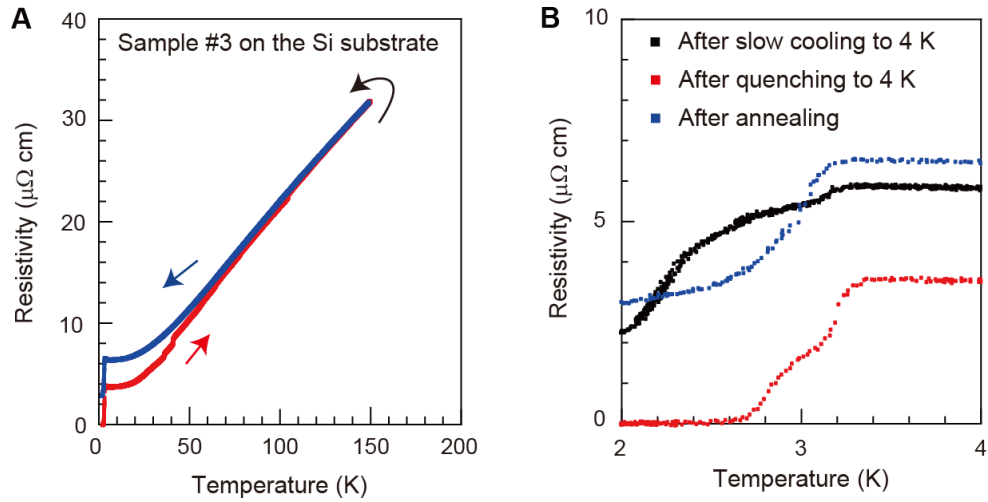

**Fig. S6. Annealing process of the quenched metastable state in the IrTe$_2$ thin plate #3 on the Si substrate.** (**A**) The temperature-resistivity profile upon heating from 2 to 150 K at a rate of $3\times10^{-3}$ K s$^{-1}$ (the red curve) and subsequent cooling to 2 K at a rate of $10^{-2}$ K s$^{-1}$ (the blue curve). Above 50 K, the red and blue curves exhibit quantitatively similar behaviors, indicating that the quenched metastable state is thermally annealed and thus relaxed into the slowly cooled state. (**B**) The temperature-resistivity profiles at low temperatures after slow cooling or quenching to 4 K and after the annealing process. The applied quenching rate is >$10^7$ K s$^{-1}$.

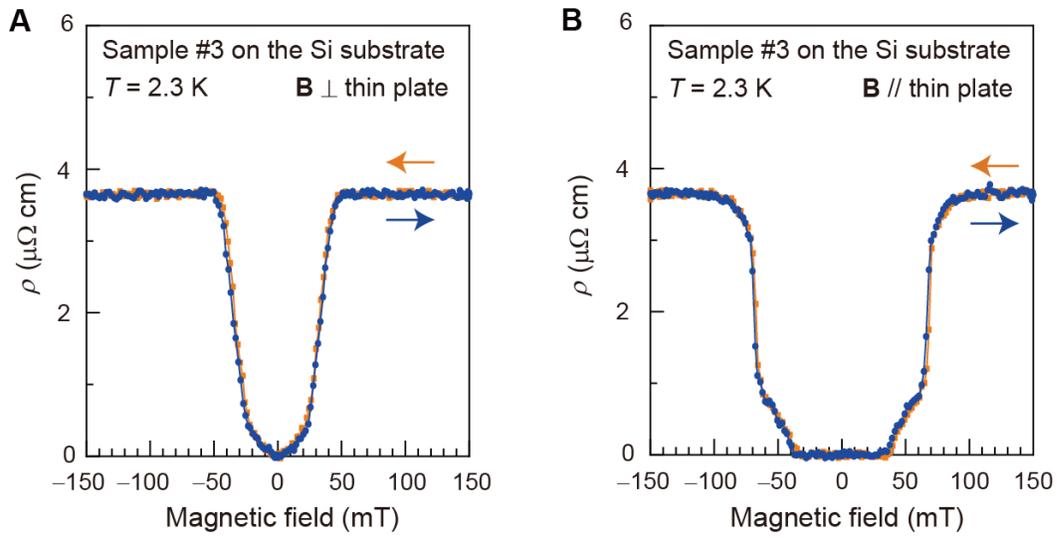

**Fig. S7. Magnetic-field effects on the emergent superconductivity in the quenched metastable states.** (**A** and **B**) The magnetic-field dependence of the resistivity with the field direction perpendicular (A) and parallel (B) to the IrTe$_2$ thin plate #3.